# ENHANCING SPARQL QUERY REWRITING FOR COMPLEX ONTOLOGY ALIGNMENTS

Anicet Lepetit Ondo[1], Laurence Capus[1] and Mamadou Bousso[2]

[1] Dep. of Computer and Software Engineering, Laval University, Quebec, G1V 0A6, Quebec, Canada

[2] Dep. Computer Science, Iba Der Thiam University, Thies, 967, Senegal

*ABSTRACT*

*SPARQL query rewriting is a fundamental mechanism for uniformly querying heterogeneous ontologies in the Linked Data Web. However, the complexity of ontology alignments, particularly rich correspondences (c : c), makes this process challenging. Existing approaches primarily focus on simple (s : s) and partially complex (s : c) alignments, thereby overlooking the challenges posed by more expressive alignments. Moreover, the intricate syntax of SPARQL presents a barrier for non-expert users seeking to fully exploit the knowledge encapsulated in ontologies. This article proposes an innovative approach for the automatic rewriting of SPARQL queries from a source ontology to a target ontology, based on a user's need expressed in natural language. It leverages the principles of equivalence transitivity as well as the advanced capabilities of large language models such as GPT-4. By integrating these elements, this approach stands out for its ability to efficiently handle complex alignments, particularly (c : c) correspondences, by fully exploiting their expressiveness. Additionally, it facilitates access to aligned ontologies for users unfamiliar with SPARQL, providing a flexible solution for querying heterogeneous data.*

## 1. INTRODUCTION

In the Linked Data Web, aligned ontologies play a crucial role in facilitating interoperability between different data sources. Thanks to these alignments, it becomes possible to reformulate queries and exploit heterogeneous knowledge coherently in an information retrieval context. However, querying them, a critical step to fully leverage the knowledge encapsulated within them, remains a significant challenge. The SPARQL language, the standard for querying RDF/OWL ontologies, imposes constraints related to its strict syntax and its dependence on the internal structure of the ontology. These characteristics limit the accessibility of this technology to non-expert users, particularly those who do not master the complex concepts of ontologies or the subtleties of the SPARQL language. This access barrier constitutes the first motivation for considering solutions that facilitate ontology use for users with varying profiles.

Meanwhile, within the same domain, the development of multiple ontologies is a common practice, but it leads to varying levels of heterogeneity due to differences in language, format, level of detail, and coverage. This high level of heterogeneity complicates interoperability between these ontologies. These disparities are mitigated through ontology alignment. Once these data sources are aligned, querying them homogenously while only knowing the vocabulary of a specific source becomes very complex. Thus, a second motivation is to offer tools capable of automatically rewriting SPARQL queries in aligned ontology contexts.

Homogeneous querying of aligned ontologies is a multifaceted problem. For a given ontology, a SPARQL query formulated in a source context must be adapted to target an aligned ontology, potentially very different in its vocabulary and structure. This process is relatively simple when it comes to simple alignments (s : s). However, it becomes increasingly complex when dealing with rich alignments involving complex expressions (c : c), where multiple relations and additional constraints must be considered **[1][2]**. Current research mainly focuses on simple (s : s) or partially complex (s : c) alignments, largely neglecting more expressive alignments**[1][2][3]**. This gap limits the exploitation of ontology alignments in contexts where the expressiveness of queries is essential to meet complex querying needs in heterogeneous environments.

In light of these barriers, our research focuses on two key issues:

- How can we enable efficient and accessible querying of ontologies in environments where users may not necessarily master SPARQL or the structure of ontologies?
- How can we ensure automatic, correct, and robust SPARQL query rewriting between aligned ontologies, especially when alignments are complex (c : c) ?

Our research aims to propose solutions to facilitate querying ontologies by non-expert users and to ensure automatic and robust SPARQL query rewriting in aligned ontology contexts, with a particular focus on complex (c : c) correspondences, which provide enhanced expressiveness, thereby optimizing the use of aligned data sources.

The major contribution of this paper lies in the development of an innovative approach to automatic SPARQL query rewriting, handling complex (c : c) alignments and leveraging equivalence transitivity to infer new implicit correspondences by linking indirectly equivalent entities, thus improving integration and consistency of inter-ontology knowledge.

To address the research problem and achieve the defined objectives, this article is structured as follows: The first section presents a state of the art on the reuse of ontology alignments, examining existing approaches and their limitations. The second section introduces the proposed approach, which integrates natural language querying of aligned ontologies, focusing on complex correspondences (c : c) and their automatic rewriting. The final section is dedicated to experimentation, where test scenarios, evaluation methods, and results are presented, followed by a discussion on the effectiveness and limitations of the proposed approach.

## 2. RELATED WORK

Linked Data on the Web relies on the use of ontologies to ensure the sharing, reuse, and interoperability of data, thus enabling a computational understanding of knowledge. In a given domain, it is common to develop multiple ontologies, leading to various levels of heterogeneity due to differences in language, format, detail, and coverage among these ontologies. Furthermore, ontology elements may present ambiguities or inconsistencies in their meanings, interpretations, or representations.

To overcome these challenges, specific tools and techniques allow for the reconciliation of these ontologies, a process known as ontology alignment. The goal of alignment is to reduce the heterogeneity issues between aligned ontologies. An alignment between two ontologies, $O$ and $O'$, is defined by a set of correspondences ⟨e, e′, r, n⟩ **[4]**. Each correspondence represents a relation r, which can either be equivalence (≡) or a subsumption relation (⊒, ⊑) between members e and e′, and n indicates the confidence threshold [0...1] of this correspondence. A member can be a simple entity of the ontology (class, data or object property, or individual) from $O$ or $O'$, or a more complex construction composed of entities and connectors.

When both members are represented in an elementary and direct manner, these are referred to as simple correspondences of type (s : s). This type of correspondence is frequently encountered in a significant portion of existing research, mainly due to its simplicity and ease of implementation. Although simple correspondences (such as (s : s) relations) are commonly used, they have significant limitations, particularly in their ability to capture the semantic richness of complex relationships within ontologies. A (s : s) correspondence only establishes a direct relationship between two similar or equivalent entities without considering the nuances or broader contexts in which these entities may be interpreted. An example of these simple correspondences, based on equivalence, extracted from the conference dataset related to the alignment between the *ekaw.owl* and *edas.owl* ontologies, is illustrated in Figure 1.

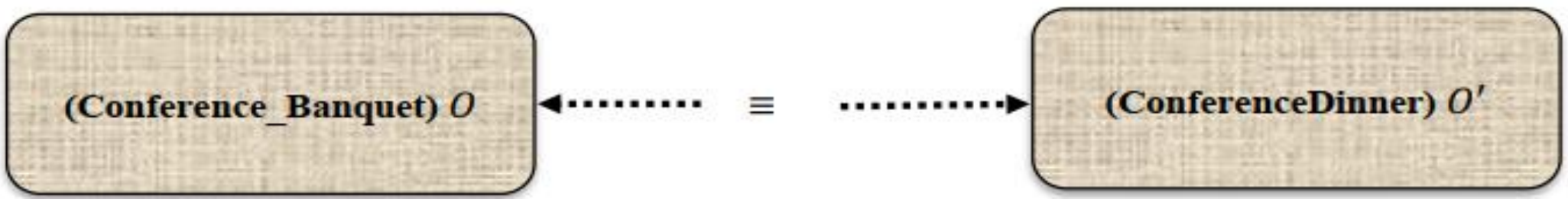

Figure 1. An Example of a Simple Alignment (s : s)

In cases where one member of the correspondence is simple while the other constitutes a complex construction, the correspondence is referred to as a complex correspondence of type (s : c) or (c : s), depending on whether the simple or complex member is on the source or target side of the correspondence. These correspondences enable the linking of a simple entity, such as an isolated concept, to a more elaborate entity that may include additional relationships or restrictions.

Alternatively, when both members of the correspondence are complex constructions i.e., entities containing multiple nested relationships or attributes the correspondence is classified as a complex correspondence of type (c : c). This type of correspondence addresses cases where multi-level relationships or hierarchical entities need to be aligned, often due to the underlying complexity of the ontology.

Figures 2 and 3 respectively illustrate these two types of correspondences based on equivalence, extracted from the same conference dataset related to the alignment between the *ekaw.owl* and *edas.owl* ontologies.

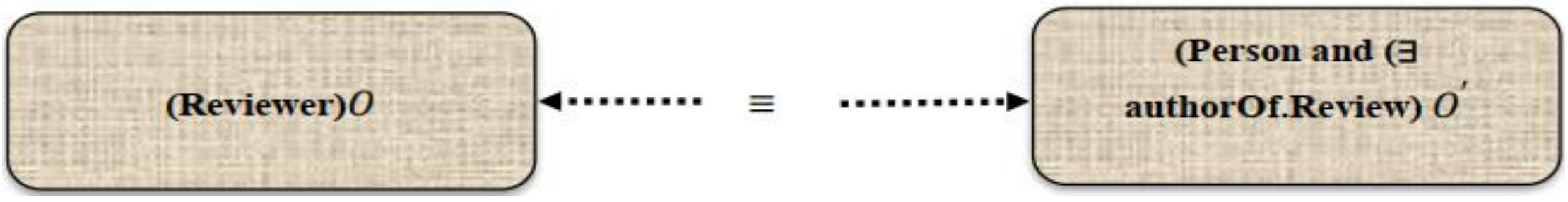

Figure 2. An Example of Complex Alignment (c : s) or (s : c)

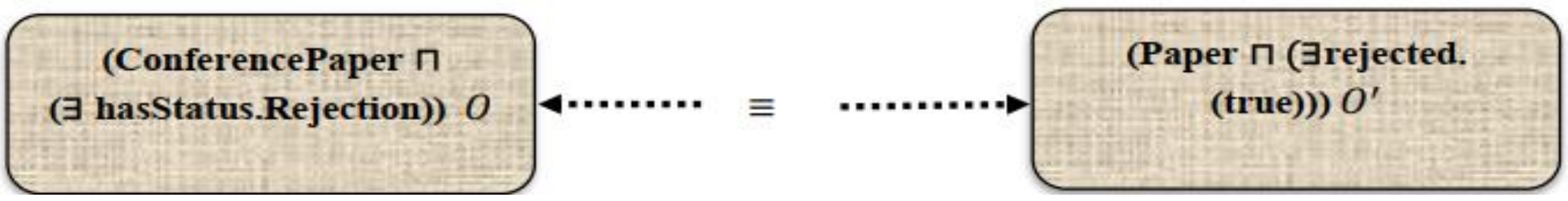

Figure 3. An Example of Complex Alignment (c : c)

To reduce disparities between data sources, several methods have been explored, including terminology methods, which address heterogeneities linked to the choice of representation format, as well as those resulting from synonymy, polysemy, words from different languages (French, English, Italian, etc.), and syntactic variations (different pronunciations, abbreviations, prefixes and suffixes, etc.) **[5][6].** The internal and external structural methods of ontologies, encompassing entities, their relationships and the properties they share, are based on the parent-child concepts defined in their respective ontologies **[5][6].** Extensional methods, on the other hand, base their reasoning on the instances (extensions of entities) of ontologies **[7][8][9]**. Some of these methods can be combined to take advantage of the benefits of each and improve alignment accuracy.

Despite significant advancements in the study of complex ontology alignments, their practical application, particularly regarding the automated rewriting of SPARQL queries in interoperable data environments, remains a largely unexplored research problem, requiring thorough investigation. Furthermore, recent contributions addressing this issue remain rare.

To address the various challenges outlined above, many research efforts aim to propose solutions to mitigate this problem. In the literature, the majority of contributions focus primarily on rewriting SPARQL queries based on simple correspondences of type (s : s), where each source entity is directly associated with an equivalent target entity. Typically, this involves substituting the IRI of an entity in the original query with the corresponding IRI in the alignment **[10]**. However, some studies address query rewriting by incorporating complex correspondences **[11][1]**, but they often limit themselves to correspondences of type (s : c). These studies propose transformation rules to reformulate SELECT-type SPARQL queries, adapting a query originally designed for one knowledge base to another while using its vocabulary. However, these approaches remain confined to complex alignments of type (s : c) and do not integrate key features, such as the management of relation transitivity, which limits their application in more complex scenarios.

In the same vein of complex alignment rewriting, a study **[2]** explores the rewriting of SPARQL queries through two distinct systems. The first focuses on rewriting triplets using simple correspondences, without the ability to combine them. In contrast, the second system, based on instances, handles complex alignments, enabling more efficient query rewriting while incorporating logical correspondences. However, these systems remain limited to correspondences of type (s : c) or (c : s), leaving the management of more complex (c : c) correspondences as an open issue.

By integrating complex correspondences based on description logic, other research **[12][13]** presented SPARQL-RW, a system capable of rewriting queries by incorporating complex correspondences based on description logic. This system handles classes, object properties, data properties, and individuals, substituting RDF models while preserving the initial variables. However, it does not support correspondences between two complex members from the source and target ontologies.

Meanwhile, Gillet et al. **[14]** propose a query pattern rewriting method adapted to query families. This method includes both simple and complex alignments but remains limited in its application to (s : c) correspondences. Additionally, Pankowski's work **[3]** offers innovative and user-friendly methods for formulating queries in ontology-based systems, facilitating the exploration of aligned data for users less familiar with complex structures. However, this approach has limitations, notably the lack of support for query rewriting in the context of more expressive (c : c) alignments and the management of relation transitivity, a crucial aspect for more complex ontological structures.

The examined works highlight the significant challenges related to SPARQL query rewriting in the context of aligned ontologies. While approaches have been developed to handle simple correspondences (s : s) and some complex correspondences (s : c), gaps remain, particularly in managing more complex correspondences (c : c) and integrating transitivity mechanisms for relations, which enhance inference and information retrieval by deducing implicit relationships from those explicitly defined, thereby enriching query responses.

These limitations highlight the pressing need for a more integrated and robust approach to query rewriting, one that can effectively navigate the richness and diversity of ontologies.

Additionally, automating this process within interoperable data environments remains an underexplored research area. Therefore, fostering future research dedicated to developing innovative methods for managing complex correspondences and transitivity is crucial to improve the querying of heterogeneous data and optimize knowledge integration across the linked data web.

## 3. SPARQL QUERY REWRITING METHODOLOGY

This section presents our comprehensive methodology for generating and rewriting queries. We begin by outlining a robust framework for query formulation, which includes guidelines for creating effective queries and understanding correspondences between ontological elements. We pursue description of our method then focuses on rewriting queries from a source ontology to their equivalent in a target ontology, ensuring semantic consistency across different structures.

### 3.1. General Structure of Correspondences Identified by the Approach

The following structure for query formulation takes into account the specificities necessary to create a robust and reproducible method for generating queries that meet the requirements of our research.

$$\pi_V(\sigma_{condition}(T_1, T_2, ..., T_n)) \quad \textbf{(1)}$$

- $\boldsymbol{\pi_V}$: It represents the projection on the set of variables V, corresponding to the elements of the SPARQL query to be extracted, with V= $\{V_1, V_2, \dots, V_m\}$ where $m$ denotesthe total number of distinct variables generated **[15]**.
- $\boldsymbol{\sigma_{condition}}$ : Represents the selection of triples that satisfy the conditions specified in the WHERE clause, if any conditions are present **[15]**.
- $\boldsymbol{T_1, T_2, \dots, T_n}$: a set of n triples, where each triple is of the form $\langle s_i, p_i, o_i \rangle$ for pour **i** ∈ {1, 2, …, n}, representing the subject, predicate, and object in the ontology, respectively **[15]**.

The representation used in formula **(1)** corresponds to the structure of the following SPARQL query:

$$SELECT \ \ ?V \ \ WHERE\{(s_1, p_1, o_1)...(s_n, p_n, o_n)\} \quad \textbf{(2)}$$

This SPARQL query rewriting approach ensures a homogeneous interrogation of heterogeneous ontologies in the Linked Data Web by leveraging natural language. Its objective is to generate the sets V and T, representing respectively a set of variables and a set of triples present in the SPARQL query of the ontology $O$ (source), and then transform them into the sets V′ and T′ corresponding to the ontology $O'$ (target) . Thus, the transformation consists of:

$$\begin{array}{c}(O):\pi_V(\sigma_{condition}(T_1, T_2, ..., T_n)) \\ \Downarrow \\ (O^{'}):\pi_{V^{'}}(\sigma_{condition}(T_1^{'}, T_2^{'}, ..., T_n^{'}))\end{array} \quad \textbf{(3)}$$

We explore three types of correspondences in our approach: simple correspondences (s : s), complex correspondences (s : c), and (c : c). However, we pay particular attention to (c : c) correspondences, which are based on equivalence relations between complex entities identified in the source and target ontologies. The general structure of correspondences to be translated into SPARQL queries for the source and target ontologies is listed in Table 1. In this table, we have documented various correspondences that may exist between entities during an ontology alignment process. These correspondences have been categorized into different patterns, allowing us to generate appropriate SPARQL queries based on the identified patterns. In this respect, we have defined several types of Patterns, including : Class Correspondence Patterns, where a concept or class in one ontology corresponds exactly to a single concept or class in another ontology; Class By Attribute Type (CAT) **[16] [17],** which involves type restrictions on the objects of a relation (property on objects); Class By Attribute Value (CAV) **[16] [17]**, which refers to value restrictions on the objects of an attribute (relation or property); and class matching patterns based on union and intersection operations.

Table 1. General Structure of Identified Correspondences

<table>
<tr><th>Types of correspondences</th><th>Global structure of identified correspondences</th><th>Pattern types</th></tr>
<tr><td>Simple Correspondences (s : s)</td><td>$\forall x(Cl_1(x) \equiv Cl_1^{'}(x))$<br>**(4)**</td><td>Class correspondance pattern</td></tr>
<tr><td rowspan="3">Complex Correspondences (s : c) or (c : s)</td><td>$\forall x(Cl_1(x) \wedge \exists y(OP(x,y) \wedge Cl_2(y)) \equiv Cl_1^{'}(x))$<br>**(5)**<br>$\forall x(Cl_1(x) \wedge \forall y(OP(x,y) \rightarrow Cl_2(y)) \equiv Cl_1^{'}(x))$<br>**(6)**</td><td>CAT correspondence pattern</td></tr>
<tr><td>$\forall x(Cl_1(x) \equiv (Cl_1^{'}(x) \wedge DP^{'}(x, value)))$<br>**(7)**</td><td>CAV correspondence Pattern</td></tr>
<tr><td>$\forall x(Cl_1(x) \equiv (Cl_1^{'}(x) \vee Cl_2^{'}(x) \vee ... \vee Cl_k^{'}(x)))$<br>**(8)**<br>$\forall x(Cl_1(x) \equiv (Cl_1^{'}(x) \wedge Cl_2^{'}(x) \wedge ... \wedge Cl_k^{'}(x)))$<br>**(9)**<br>$k$ represents the total number of classes in the target ontology $O^{'}$ that are equivalent to class $Cl_1$ from the source ontology $O$ through a logical disjunction or conjunction</td><td>CU or CI correspondence pattern</td></tr>
<tr><td>Complex Correspondences (c : c)</td><td>$F \equiv F^{'}$<br>**(10)**<br>The formulas $F$ and $F^{'}$ can correspond to complex members of an ontology, defined as constructs of entities and connectors, identified in this table under the category of (c : s) or (s : c) type correspondences.</td><td>No specific type, the SPARQL query is constructed based on the identified complex correspondences</td></tr>
</table>

Let :

- $Cl, OP, DP \in E$ $and$ $Cl', OP', DP' \in E'$, where $E$ $and$ $E'$ represent an entity (classes and properties) of the source ontology $O$ and the target ontology $O'$, respectively.
- $Cl = <Cl_1, Cl_2, \ldots, Cl_n>$ be the classes of a source ontology, and $Cl' = <Cl'_1, Cl'_2, \ldots, Cl'_m>$ be the classes of a target ontology, where **n** and **m** represent the total number of classes in the source ontology and the target ontology, respectively.
- The object properties of a source ontology are denoted as $OP = <OP_1, OP_2, \ldots, OP_n>$,and the object properties of the target ontology are denoted as $OP' = <OP'_1, OP'_2, \ldots, OP'_m>$, where **n** and **m** represent the total number of object properties in the source ontology and the target ontology, respectively.
- $DP = <DP_1, DP_2, \ldots, DP_n>$ represents the data properties of our source ontology, and $DP' = <DP'_1, DP'_2, \ldots, DP'_m>$, represents those of the target ontology, where **n** and **m** correspond respectively to the total number of data properties in the source ontology and the target ontology.

The following abbreviations are also to be considered:

- **CAV**: Class by Attribute Value
- **CAT**: Class by Attribute Type
- **CU**: Class Union
- **CI**: Class Intersection

## 3.2. Method Adopted

By leveraging the aligned ontology, our approach consists of rewriting a query formulated in the source ontology into its equivalent in the target ontology. The process includes several essential steps, made possible through Owlready2, a Python library that allows the manipulation of ontologies in OWL format. This library is essential for ontology management, analysis, and alignment, as well as for integrating prompts with the LLM GPT-4, which, based on a textual input, guides the model generation according to the provided instructions.

Our approach begins with the introduction of a question provided by a user, which is then subjected to a normalization phase to standardize terms and facilitate their processing in the context of extracting relevant concepts.

Next, a Python function is developed to synchronize, beforehand, the Pellet reasoner with the aligned ontology. This reasoner is crucial in our approach because it infers implicit axioms by examining explicit axioms and applying logical rules to deduce additional knowledge, even if it is not explicitly mentioned in the aligned data. Subsequently, this function analyzes the source ontology's entities and verifies their equivalences with those in the target ontology. When an entity corresponds to a complex formula (conjunctions, disjunctions, existential or universal restrictions), the function extracts its components, processes them, and assembles them to create a readable representation. The identified correspondences are added to a key-value dictionary, where the key represents the subgraph from the source ontology, and the value represents its equivalent in the target ontology. The function also identifies simple equivalences, where one entity is directly equivalent to another, and incorporates them into the same dictionary. This initial extraction thus detects direct correspondences (s : s) and (s : c) or (c : s) present in the ontology, as well as those inferred by the reasoner.

The simple and complex correspondences (s : c) or (c : s) identified in the previous phase undergo further evaluation to infer new correspondences that were not previously detected. This includes equivalences based on more complex constructions, composed of entities and connectors in both the source and target ontologies, leading to (c : c) correspondences. This process enriches the knowledge base and improves ontology alignment accuracy, particularly through the application of equivalence transitivity rules. The newly generated complex correspondences are then added to the dictionary to enhance the alignment.

Once all correspondences are identified and stored in the dictionary, the approach uses the normalized phrase to search for the corresponding key. The GPT-4 prompt provides the context or framework in which the model should generate a response. This prompt associates phrase expressions with dictionary keys, establishing precise links with relevant triples. This method accounts for synonyms, complex structures, and syntactic and semantic relationships, ensuring an accurate extraction of dictionary key correspondences adapted to the context of the phrase.

Finally, SPARQL queries for the target ontology are generated based on rules derived from identified patterns, using the equivalent value of the key in the dictionary. Since a key in a programming dictionary must be unique, in our approach, when a source ontology entity identified as a dictionary key is associated with multiple equivalent values (which may be atomic entities or a formula) in the target ontology, a nested data structure, specifically a list, is used to associate these values with the key. Thus, SPARQL query rewriting is applied to each list composed of the elements of the dictionary key, representing the relevant subgraph of the source ontology, as well as to each list of dictionary values corresponding to the equivalent subgraph in the target ontology.

## 4. EXPERIMENTS AND RESULTS

This section represents a key stage in our approach. It marks the transition from theoretical design to experimental validation, where the proposed method for integrating natural language into the querying of complex aligned ontologies is put to the test. In this context, the experiments aim to assess the relevance, efficiency, and robustness of our solution in addressing the challenges posed by complex alignments and the needs of non-expert users. We first present the experimental protocol, which provides a detailed description of the dataset used. Then, an in-depth analysis of the obtained results allows us to identify the strengths and limitations of the proposed solution.

### 4.1. Manually Enrich the Dataset

The dataset is a crucial element for testing our approach. However, the available datasets did not allow us to validate all the models of complex correspondences that our method must evaluate. Indeed, they only support certain types of correspondences, particularly those of the form (s : s), which link two simple (or atomic) entities between a source ontology and a target ontology, as well as (s : c) correspondences.

To obtain a dataset capable of inferring more complex correspondences of the type (c : c), which link two propositions (or formulas), we manually enriched the complex alignment between the two ontologies (ekaw.owl and edas.owl) present in the Conference dataset, focusing particularly on the **TBOX** part. We used an existing alignment in this dataset, which we then enriched to infer equivalence transitivity rules, thereby linking two formulas from a source ontology and a target ontology according to the following two reasoning patterns:

**Reasoning 1** : $R(E, F_1^{'}) \wedge R(E, F_1) \Rightarrow R(F_1, F_1^{'})$ **(11)**

**Reasoning 2** : $R(E, E^{'}) \wedge R(E, F_1) \wedge R(E^{'}, F_1^{'}) \Rightarrow R(F_1, F_1^{'})$ **(12)**

With :

- $R$, an equivalence relation
- $F_1$, a complex construction (composition of several entities) from the source ontology $O$
- $F_1^{'}$, a complex construction (composition of several entities) from the target ontology $O'$

To achieve this, we relied on our experience with concepts related to scientific conferences as well as specialized forums to identify more complex relationships. Our procedure consisted of starting with two equivalent entities between a source ontology and a target ontology, then searching for an equivalent proposition for each of these entities. This proposition is then translated into a formula and added to the existing dataset (i.e., the alignment between the ekaw and edas ontologies). To further refine the meaning of the formulas to be implemented, we also added new classes and properties, thus expanding the range of possible complex correspondences. Figures 4 and 5 present examples of equivalent formulas that we can associate with each atomic entity and also illustrate a small reasoning mechanism that could be used to apply equivalence transitivity.

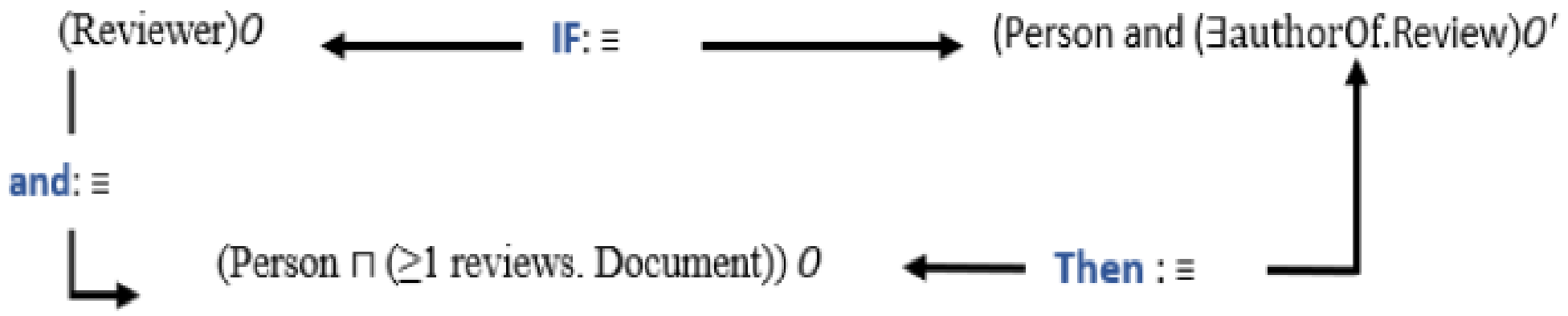

Figure 4. Example of Inference Based on Reasoning 1

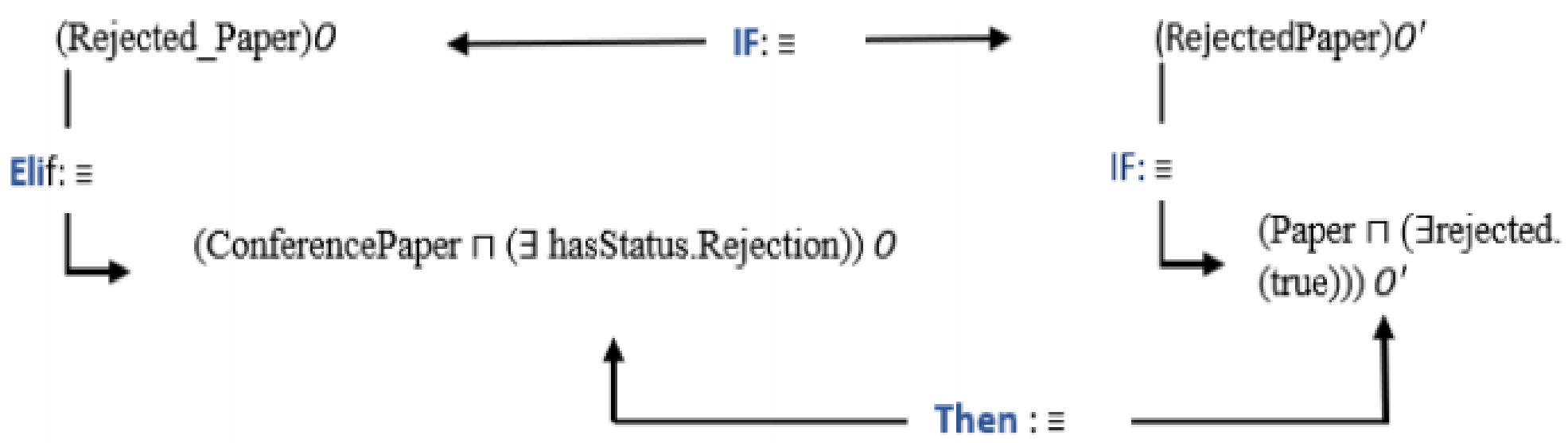

Figure 5. Example of Inference Based on Reasoning 2

Figures 6 and 7 below illustrate the structure of the dataset used to validate our approach. This dataset establishes correspondences between two ontologies from the conference domain, which have been previously enriched. Figure 6 presents examples of simple correspondences, while Figure 7 illustrates complex correspondences of the (s : c) type, as well as the process for identifying more expressive correspondences of the (c : c) type. In particular, Figure 7 shows how certain relationships extend the scope of correspondences towards more complex forms. For example:

$Accepted_Paper \equiv AcceptedPaper,$
$Accepted_Paper \equiv ConferencePaper\ and\ (hasDecision\ some\ Acceptance),$

$$AcceptedPaper \equiv Paper\ and\ (accepted\ value\ true)$$

Based on the three previous reasonings, we can infer a complex correspondence of the (c : c) type, as illustrated below :

$$ConferencePaper\ and\ (hasDecision\ some\ Acceptance) \equiv Paper\ and\ (accepted\ value\ true)$$

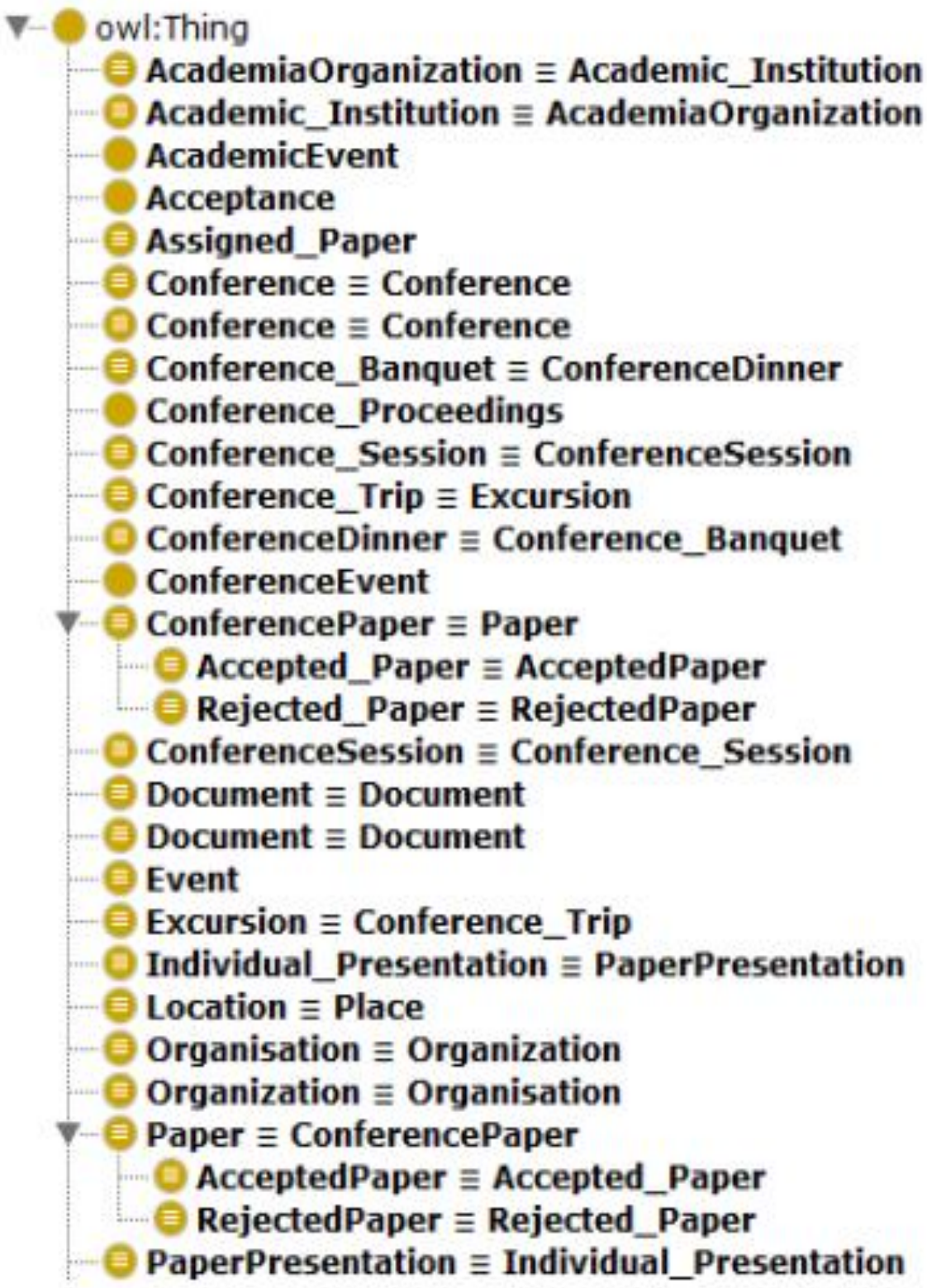

Figure 6. Types of Simple Correspondences (s : s)

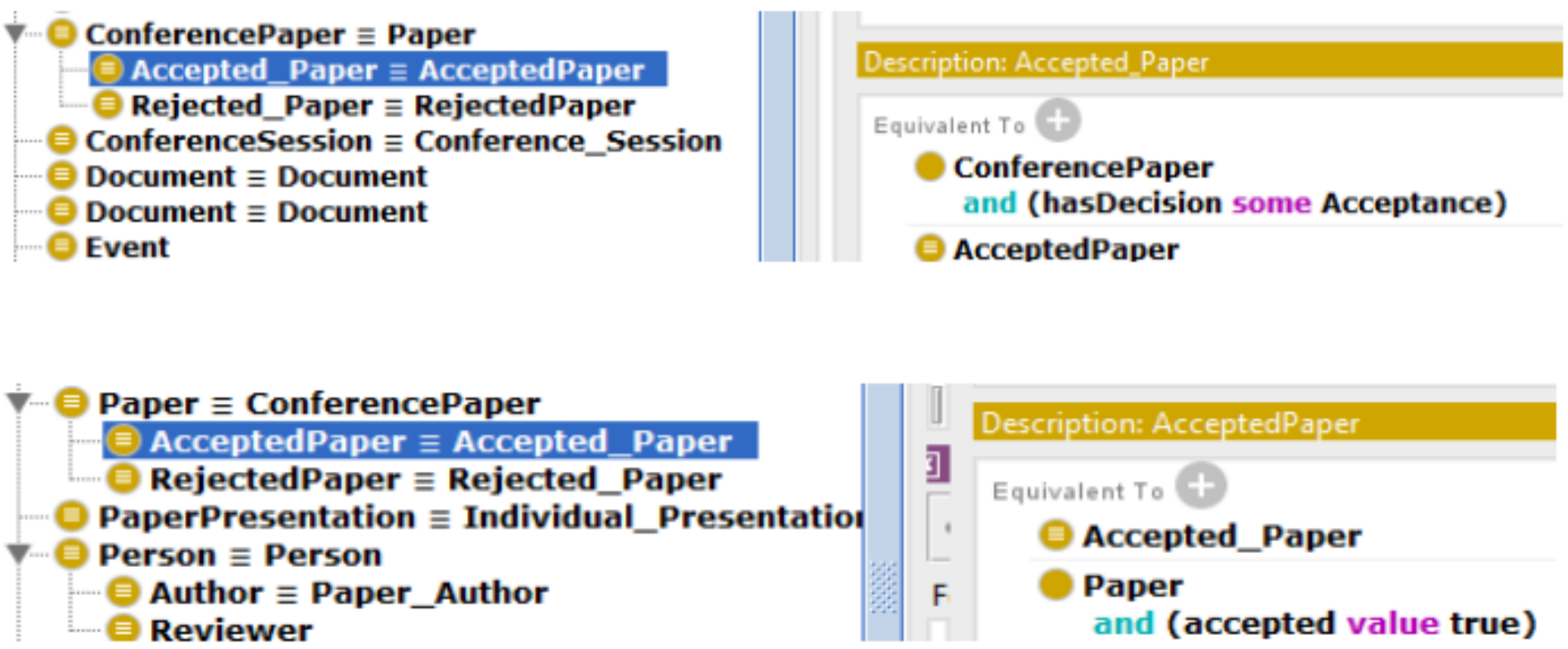

Figure 7. Characterization of Complex Correspondences of Type (s : c) with Potential Inference of Relationships of Type (c : c)

## 4.2. Tests Conducted

The tests conducted in this study aim to evaluate the effectiveness of SPARQL query rewriting by leveraging three types of correspondences defined in our aligned ontology. To implement our approach, we used the Google Colab environment, which facilitated the efficient execution of Python-based code, particularly by exploiting the powerful capabilities of the Owlready2 library and the GPT-4 model.

Our methodology consists of expressing a user need in natural language, translating it into a subgraph derived from the ontology correspondences, and then analyzing the generated SPARQL queries. This approach ensures a systematic evaluation of how the rewritten queries align with the expected semantic correspondences. Below, we present illustrative examples of the obtained results :

### a. Simple correspondences (s : s): Class correspondence pattern

This is the most common case in many studies, where an entity in a source ontology is equivalent to an entity in a target ontology. Let us consider this type of correspondence:

$$\forall x(Cl_1(x) \equiv Cl_1^{'}(x))$$

We will automatically rewrite:

$$T_1(O) = <s_1, p_1, o_1>, \text{with} \begin{cases} s_1 \in V \\ p_1(rdf:type) \\ o_1 \in Cl \end{cases} \longmapsto T_1^{'}(O^{'}) = <s_1^{'}, p_1^{'}, o_1^{'}>, \text{with} \begin{cases} s_1^{'} \in V^{'} \\ p_1^{'}(rdf:type) \\ o_1^{'} \in Cl^{'} \end{cases} \quad \textbf{(13)}$$

Let us consider the four reformulations below, illustrating the same need to test our simple correspondence type (s: s):

- *Formulation1: Can you tell me about the various conference receptions?*
- *Formulation2: What are the various banquets held during the conference?*
- *Formulation3: Can you provide information about the different banquets at the conference?*
- *Formulation4: Could you list the different conference banquets?*

Figure 8 below presents the result obtained based on the four evaluated need variations for our class correspondence model:

```
Please formulate your request  : Could you list the different conference banquets ?
Onto source = ['Conference_Banquet']
Onto cible = ['ConferenceDinner']
Appropriate SPARQL query on the source side              | Automatically generated query on the target side
-------------------------------------------------------- | -------------------------------------------------
SELECT ?variable                                         | SELECT ?variable
WHERE {                                                  | WHERE {
  ?variable rdf:type source :Conference_Banquet .        |   ?variable rdf:type target :ConferenceDinner .
  }                                                      |   }
```

Figure 8. SPARQL query rewriting in the Case of Simple Alignment (s : s)

## b. Complex correspondences (s: c) or (c: s): CU pattern

Here, we present cases where an element from a source ontology is associated with multiple elements in a target ontology, and vice versa. For example, consider this type of correspondence: $\forall x(Cl_1(x) \equiv (Cl_1^{'}(x) \vee Cl_2^{'}(x) \vee ... \vee Cl_k^{'}(x)))$

We will automatically rewrite:

$$T_1(O) = <s_1, p_1, o_1>, \text{with} \begin{cases} s_1 \in V \\ p_1(rdf:type) \\ o_1 \in Cl \end{cases} \quad (14)$$

$$T_1^{'}(O^{'}) \vee T_2^{'}(O^{'}) \vee ... \vee T_n^{'}(O^{'}) = <s_1^{'}, p_1^{'}, o_1^{'}> \vee <s_2^{'}, p_2^{'}, o_2^{'}> \vee ... \vee <s_n^{'}, p_n^{'}, o_n^{'}>, \text{with}$$

$$\begin{cases} s_1^{'}, s_2^{'}, ..., s_n^{'} \in V^{'} \\ p_1^{'}, p_2^{'}, ..., p_n^{'}(rdf:type) \\ o_1^{'}, o_2^{'}, ..., o_n^{'} \in Cl^{'} \end{cases}$$

Figure 9 below presents the result for the following correspondence:

$$\forall x(Event(x) \equiv (Conference(x) \vee ConferenceEvent(x) \vee ConferenceSession(x)))$$

Imagine that a user reformulates the following questions to query an aligned source:

- *Formulation1: What kinds of events are there?*
- *Formulation2: Can you tell me about the different event?*
- *Formulation3: What types of events exist?*

The query generated on the source and target ontologies, based on the defined variations, is as follows:

```
Please formulate your request  : What types of events exist?
Onto source = ['Event']
Onto cible = ['Conference', 'ConferenceEvent', 'ConferenceSession']
Appropriate SPARQL query on the source side                  | Automatically generated query on the target side
------------------------------------------------------------ | -----------------------------------------------------
SELECT ?variable                                             | SELECT ?variable
WHERE {                                                      | WHERE {
  ?variable rdf:type source :Event .                         |   ?variable rdf:type target :Conference .
  }                                                          |   }
                                                             |   UNION
                                                             |   {
                                                             |   ?variable rdf:type target :ConferenceEvent .
                                                             |   }
                                                             |   UNION
                                                             |   {
                                                             |   ?variable rdf:type target :ConferenceSession .
                                                             |   }
```

Figure 9. SPARQL query rewriting in a case of complex alignment (s : c) or (c : s)

## c. Complex correspondences (c : c)

This category concerns cases where a complex construction of entities and connectors from a source ontology is equivalent to another complex construction from a target ontology. For example, consider this type of correspondence:

$$\forall x((Cl_1(x) \vee (Cl_2(x) \wedge \exists y(OP(x,y) \wedge Cl_3(y)))) \equiv (Cl_1^{'}(x) \vee (Cl_2^{'}(x) \wedge DP^{'}(x, value))))$$

We will need to automatically rewrite:

$$T_1(O) \vee (T_2(O), T_3(O)) = <s_1, p_1, o_1> \vee (<s_2, p_2, o_2>, <s_3, p_3, o_3>), \text{with} \begin{cases} s_1, s_2, s_3, o_3 \in V \\ p_1, p_2 (rdf:type) \\ o_1, o_2 \in Cl,\ p_3 \in OP \end{cases}$$

$$\downarrow \qquad \textbf{(15)}$$

$$T_1^{'}(O^{'}) \vee (T_2^{'}(O^{'}), T_3^{'}(O^{'})) = <s_1^{'}, p_1^{'}, o_1^{'}> \vee (<s_2^{'}, p_2^{'}, o_2^{'}>, <s_3^{'}, p_3^{'}, o_3^{'}>), \text{with} \begin{cases} s_1^{'}, s_2^{'}, s_3^{'} \in V^{'} \\ p_1^{'}, p_2^{'} (rdf:type), p_3^{'} \in DP^{'} \\ o_1^{'}, o_2^{'} \in Cl^{'}, o_3^{'} (literal) \end{cases}$$

Figure 10 below presents the result for the following correspondence:

$$\forall x((Accepted_Paper(x) \vee (ConferencePaper(x) \wedge \exists y(hasDecision(x,y) \wedge Acceptance(y)))) \equiv (AcceptedPaper(x) \vee (Paper(x) \wedge accepted(x, true))))$$

Suppose a user reformulates the following questions to query an aligned source:

- *Formulation1: Which conference papers have their acceptance status decided?*

- *Formulation2: Can you list the conference papers that have received a decision about acceptance ?*
- *Formulation3: Which papers in the conference have been decided upon in terms of acceptance ?*
- *Formulation4: Which conference papers have been reviewed and accepted?*

The query generated on the source and target ontologies, based on the defined variations, is as follows:

```
Please formulate your request  : Can you list the conference papers that have received a decision about acceptance ?
Onto source = ['ConferencePaper,hasDecision,Acceptance', 'Accepted_Paper']
Onto cible = ['AcceptedPaper', "Paper,accepted,value'true'"]
Appropriate SPARQL query on the source side                | Automatically generated query on the target side
---------------------------------------------------------- | ------------------------------------------------------
SELECT ?variable                                           | SELECT ?variable
WHERE {                                                    | WHERE {
  ?variable rdf:type source :Accepted_Paper .              |   ?variable rdf:type target :AcceptedPaper .
  }                                                        |   }
  UNION                                                    |   UNION
  {                                                        |   {
  ?variable rdf:type source :ConferencePaper .             |   ?variable rdf:type target :Paper .
  ?variable source:hasDecision ?Acceptance .               |   ?variable target:accepted 'true' .
  }                                                        |   }
```

Figure 10. SPARQL query rewriting in the case of complex alignment (c : c)

To evaluate our approach, we tested the alignments generated by our enriched dataset, based on the Ekaw-edas ontologies from the Conference dataset. To measure the versatility of our method, we also applied it to another ontology from the same dataset (alignment between conference.owl and confOf.owl). The results obtained show that, regardless of the type of correspondence, whether direct or indirect, between the aligned ontologies, we succeed in rewriting the queries of the subgraph from the source ontology into their equivalents in the target ontology, according to the patterns defined in Table 1. Several subgraphs were introduced to validate our approach. To present our results, we limit ourselves to one example per type of correspondence, using subgraphs from the source ontology and their found equivalents in the target ontology, as follows :

- **Simple Correspondence (s : s) :**

$$\forall x(Conference_Banquet(x) \equiv ConferenceDinner(x))$$

- **Complex Correspondences (s : c) or (c : s) :**

$$\forall x(Event(x) \equiv (Conference(x) \vee ConferenceEvent(x) \vee ConferenceSession(x)))$$

- **Complex Correspondences (c : c) :**

$$\forall x((Accepted_Paper(x) \vee (ConferencePaper(x) \wedge \exists y(hasDecision(x, y) \wedge Acceptance(y)))) \equiv (AcceptedPaper(x) \vee (Paper(x) \wedge accepted(x, true))))$$

The results we obtained during the tests are satisfactory, and the method presented in this article adapts effectively to any aligned dataset, incorporating both simple and complex correspondences.

## CONCLUSION AND PERSPECTIVES

A concise review of the literature in this field reveals that current research on our topic has made little progress and remains scarce. Moreover, it does not address complex correspondences (c: c), which are essential for fully leveraging the expressiveness offered by ontology alignments. Furthermore, mechanisms such as the transitivity of relations, which enable the inference of new implicit and complex correspondences, are absent from these approaches. Likewise, the use of natural language to automatically generate complex alignments is largely underexplored, even though existing approaches struggle with the semantic variations in user queries when extracting information from ontologies. Given the power of large language models, such as GPT-4, which rely on the attention mechanism and multi-head attention, it is crucial for research in this field to fully exploit their potential **[18]**. This is precisely what constitutes the strength and originality of our work in this article.

The objective of this work was to rewrite a SPARQL query from a source ontology to its equivalent in a target ontology within the framework of a complex ontology alignment. Natural language was integrated as the preferred means to enable universal users to explore these data sources, with particular attention given to more expressive alignments of type (c : c). Our approach explores each type of alignment presented in Table 1, carefully considering the existing restrictions in the ontologies (universal and existential quantifiers).

The initial experiments with this approach were limited to introducing a subgraph of the source ontology, followed by evaluating the SPARQL queries generated from the equivalent values in the target ontology, relying on prior knowledge of the structure of the aligned ontologies and the identified patterns. Subsequently, we introduced a module that automates the generation of the subgraph based on any question posed by a user, provided that it is deemed relevant to the targeted ontology.

Although our current approach is designed to adapt to different aligned ontologies, its main objective is to offer users even without prior knowledge of the ontology structure or SPARQL queries the ability to explore multiple aligned data sources. To achieve this goal, we leverage advanced natural language processing (NLP) techniques to analyze the user's question and extract both its intent and key entities. The integration of large language models (LLMs), such as GPT-4, has enabled us to accurately map a user's formulated question to the corresponding subgraph in the aligned ontology.

This research offers several key contributions:

- A reflection on querying ontologies in natural language, enabling a more intuitive and accessible interaction with knowledge bases.
- An in-depth exploration of complex alignments, often overlooked by existing approaches, and their integration into practical querying scenarios.
- An innovative approach to automatic SPARQL query rewriting, handling complex (c : c) alignments and leveraging equivalence transitivity to infer new implicit correspondences by linking indirectly equivalent entities, thus improving integration and consistency of inter-ontology knowledge. Additionally, by integrating large language models (LLMs) such as GPT-4, this approach provides a deeper understanding of user needs in relation to the structured information in aligned data sources.

- A flexible and robust solution tailored to heterogeneous environments, addressing the growing need for data integration and interoperability on the Linked Data Web.

The approach successfully identifies all types of correspondences present in different aligned ontologies, as illustrated in Figures 8, 9, and 10, while also adapting more effectively to syntactic and semantic variations of the same user-expressed need. Despite these advancements, it is important to note that, for now, its functionality primarily relies on correspondences based on equivalence relations.

Our future research aims to extend this approach beyond equivalence relations between entities by incorporating mechanisms to also address needs related to subsumption relations. Furthermore, we plan to test the system on large and complex datasets to evaluate its robustness. Finally, we intend to develop a user-friendly interface, such as a chatbot, to experiment with this approach in a real-world environment.

**AUTHORS**

Anicet Lepetit ONDO is a computer science researcher and Ph.D. candidate at Laval University, specializing in formal ontologies. His research focuses on improving ontology integration methods, particularly SPARQL query rewriting in heterogeneous and aligned environments. He has a background in computer science education and is the author of several articles on querying formal ontologies.

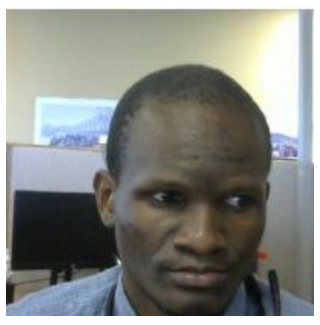

Dr. Laurence Capus is a professor of computer science at Laval university, specializing in the application of artificial intelligence to educational systems. Her work focuses on knowledge engineering, a critical aspect of artificial intelligence that involves capturing and representing expert knowledge in computer systems. She has contributed in developing intelligent tutoring systems and adaptive learning platforms that personalize education for individual students.

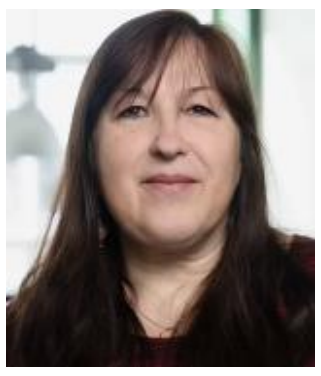

Dr. Mamadou Bousso, an Associate Professor at the University of Thiès and an Adjunct Professor at Laval University, is a researcher and lecturer in computer science specializing in complex systems modeling. His career has led him to work on interdisciplinary issues involving economists, agronomists, computer scientists, as well as healthcare professionals and administrators, both in France and Africa. This immersion in multidisciplinary teams has allowed him to strengthen his expertise in complex systems modeling and develop a strong motivation to carry out academic and research projects, particularly in the field of the semantic web. Currently, he leads the master's program in Data Science and Artificial Intelligence and supervises several research projects in artificial intelligence applied to healthcare.

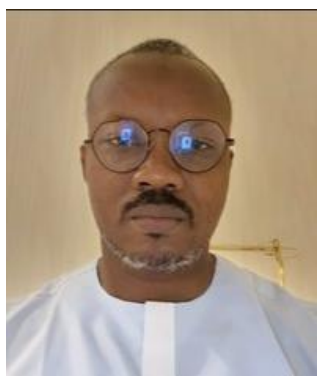